\documentclass{article}
\usepackage{frascatiphys_fnal}
\begin{document}
\title{ 
  TOP MASS MEASUREMENT AT CDF
}
\author{
  Kostas KORDAS\\
  {\em LNF-INFN, Frascati (RM) 0044, Italy} \\
  (for the CDF Collaboration)
}
\maketitle
\baselineskip=11.6pt
\begin{abstract}
 We report on recent measurements of the top quark mass using 
$t\bar{t}$ candidate events selected in $\simeq 320$ pb$^{-1}$ of data from
the "Run II" operation period of the Tevatron $p\bar{p}$ collider.
More emphasis is given on the best single measurement to date 
($M_{top} = 173.5^{+3.9}_{-3.8}$ GeV/c$^2$), provided by CDF using the
"lepton plus jets" channel, where one $W$ decays to a lepton-neutrino
pair and the other into quarks
(top quarks decay to $Wb$ almost 100\% of the time).
\end{abstract}
\baselineskip=14pt
\section{Introduction}
The top quark is the heaviest fundamental particle of the Standard Model (SM): 
about 35 times larger than the next heaviest quark ($b$) in the theory. 
A precise measurement of its' mass ($M_{top}$) is important 
for SM's prediction power, because top quark loops affect many 
Electro Weak (EW) observables. 
The top mass is also linked with the mass of the $W$ and the Higgs bosons 
via radiative corrections; a fact exploited by EW fits which use measured 
observables to simultaneously constrain the three masses assuming the 
Standard Model\cite{LEPEW2005}.
Since the Higgs eludes experimental observation, a precise measurement of 
the top quark and the $W$ masses serve as a constraint to the Higgs mass.  

 Top quarks were first observed by the 
CDF and D{\O} collaborations in 1995\cite{top_discovery},
in events produced at $p\bar{p}$ collisions of  $\sqrt{s} = 1.8$~TeV 
during the ``Run I'' operation of the Fermilab Tevatron collider.
The two experiments combined top mass measurements to find 
$M_{top} = 178.0 \pm 4.3$ GeV/c$^2$\cite{Mtop_World_Run1}.
 In the current mode of operations (``Run II'') the Tevatron provides 
$p\bar{p}$ collisions of $\sqrt{s} = 1.96$~TeV. At the time of this 
conference, it has delivered 1.3 fb$^{-1}$ of integrated luminosity,
with about 1 fb$^{-1}$ recorded by each experiment (almost ten times 
the Run I data sample). 
The Tevatron plans to deliver 4 to 8 fb$^{-1}$ per experiment, 
which aim to measure the top mass with an accuracy of $\sim 2$ GeV/c$^2$ each.

 We present here four recent $M_{top}$ measurements at CDF, 
obtained with datasets of 320 to 360 pb$^{-1}$.
More emphasis is given on a novel method to attack the dominant 
systematic uncertainty of the measurement\cite{CDF_ljets_Mtop}.

\section{Top mass reconstruction}

\subsection{Event selection}
Top quarks are mostly pair produced at the Tevatron, via
$q\bar{q}$ annihilation ($\sim~85\%$) or gluon-gluon fusion, 
and they decay $\sim 100\%$ of the time to $W b$,
with the $b$ quark hadronizing into a jet ($j$) of particles.
Subsequently, $W$'s decay to quarks ($q_1 \, \bar{q_2}$) or to a 
lepton-neutrino ($\ell \nu$) pair and their decay mode sets the 
characteristics of the $t\bar{t}$ event. 
The most precise measurements are obtained in
i) the {\em lepton plus jets} channel (``$\ell + jets$''), where 
$t\bar{t} \to W^+ \, b \, W^- \, \bar{b} \to b \, \bar{b} \, \ell \, \nu \, q_1 \, \bar{q_2}$ (30\% of 
$t\bar{t}$ decays, when $\ell \equiv e$ or $\mu$), or 
ii) the {\em dilepton} channel, where 
$t\bar{t} \to W^+ \, b \, W^- \, \bar{b} \to  b \, \bar{b} \, \ell^{\pm}_1 \, \ell^{\mp}_2 \, \nu \, \bar{\nu}$ (5\% of the time, when $\ell \equiv e$ or $\mu$). 

We select events containing a well identified and isolated 
electron or muon, large missing transverse energy from the neutrino(s) produced in the 
$W$ decay(s), and jets with a multiplicity depending on the decay channel. 
All reconstructed objects are required to have high transverse energy 
(e.g., $E_T > 15$ GeV)\cite{CDF_ljets_Mtop}. 
In the dilepton channel, a second lepton is also requested; most commonly an 
$e$ or $\mu$ candidate again, but sometimes we just ask for a well 
isolated and energetic track in order to recover some of 
the $\tau$ decays (``lepton plus track'' selection)\cite{CDF_DIL-LTRK_xsec}.
Dilepton and $\ell + jets$ samples are constructed to be mutually exclusive.

\subsection{Measurement challenges}
To reduce the combinatorial and physics background in the 
$\ell + jets$ channel, it is common to request at least one $b$-tagged jet, 
i.e., a jet identified as originating from a $b$ quark, typically via the 
presence of a secondary vertex. 
Thus, if at least one jet is $b$-tagged, the possible jet-parton assignments 
are reduced from 12 to 6, and to just two if both $b$ jets are tagged 
(the two $W$ daughter jets are interchangeable in the reconstruction).

 The largest systematic uncertainty in the $M_{top}$ 
measurement is due to the uncertainty on the Jet Energy Scale (JES, i.e., 
the jet-to-parton energy translation). 
The original parton energy is estimated by correcting the jet for instrumental,
radiation and fragmentation effects, with a $\sim 3\%$ uncertainty 
for high energy jets \cite{CDF_JES}.
The novel top mass measurement we report here, uses the hadronic $W$ decays 
in the $t\bar{t}$ data sample itself to further constrain the JES.

 Even though statistically limited, the dilepton channel has a higher signal to background ratio due to the 
second lepton. No $b$ tagging is then required. The two highest energy 
jets are assumed to be $b$ jets and we are left with just two possible 
jet-parton assignments. 
Nevertheless, the measured missing energy is due to two neutrinos 
and we are faced with an under-constrained problem. 
We thus iterate over more assumptions about the event topology
compared to the  over-constrained case of the lepton plus jets channel. 
This complication results in a larger statistical uncertainty in this channel, 
but the smaller number of jets yields a smaller JES contribution to 
the systematic uncertainty.

\subsection{Measurement methods}
\label{Sec:Mtop_methods}
 In template methods, we impose kinematic constraints on the event 
according to the $t\bar{t}$ decay hypothesis and
for each possible topological configuration we compute an event $\chi^2$
which takes into account the detector resolution 
and the $W$ and top decay widths.
We get the most probable reconstructed top mass per event and 
we compare the distribution with similarly obtained ``template'' p.d.f's
from simulated background and signal events with known $M_{top}$.
A likelihood minimization yields the $M_{top}$ which best describes the
measured distribution as an admixture of $t\bar{t}$ and background events.

 In matrix-element methods, for each $t\bar{t}$ candidate event we
calculate a likelihood, as a function of $M_{top}$, which is 
the differential probability that the measured quantities 
(4-momenta of reconstructed objects) correspond to a  
$t\bar{t}$ signal event. The likelihood is the convolution of the 
leading-order (LO) $t\bar{t}$ matrix element and detector 
resolution functions, integrated over all possible configurations.
Similarly, we can also compute a likelihood for each event being background
(e.g., $W$ plus jets in the $\ell + jets$ case, or Drell-Yan for dilepton)
and sum the likelihoods according to the relative abundance of each 
contribution. 
The product of the individual event likelihoods forms the joint likelihood 
of the data sample, which is fitted to yield the top mass measured from the sample.

\section{Top mass measurements}
\subsection{The best single measurement}
%
The most precise measurement is obtained with a 2-Dimensional template analysis
of 318 pb$^{-1}$ of $\ell + jets$ data\cite{CDF_ljets_Mtop}.
This analysis uses the dijet mass from the in-situ $W \to j j$ decays to constrain 
the jet energy scale.  
Since the dijet mass  is largely insensitive to the true top mass, we create separate 
templates of the reconstructed top and $W$ masses in simulated 
events as a function of the true top mass and JES, respectively, and
compare them with the data distributions.

In order to improve the statistical power of the method, four mutually  exclusive 
subsamples are used, defined by the number of $b$-tagged jets and the 
jet $E_T$ cuts. Sample ``1-tag(T)'' has one $b$-tag and $E_T > 15$ GeV for all jets, 
whereas ``1-tag(L)'' relaxes one jet's requirements to $8 < E_T < 15$ GeV.
 In each subsample
we select the most ``reasonable'' $t\bar{t}$ candidates by applying a cut on 
the $\chi^2$ mentioned in Section \ref{Sec:Mtop_methods}.
Figure 1 shows the reconstructed top mass in the data, with the best 
fits from the simulated templates overlaid. Similar plots are obtained for the dijet mass.
\begin{figure}[t]
 \vspace{7.5cm}
 \includegraphics{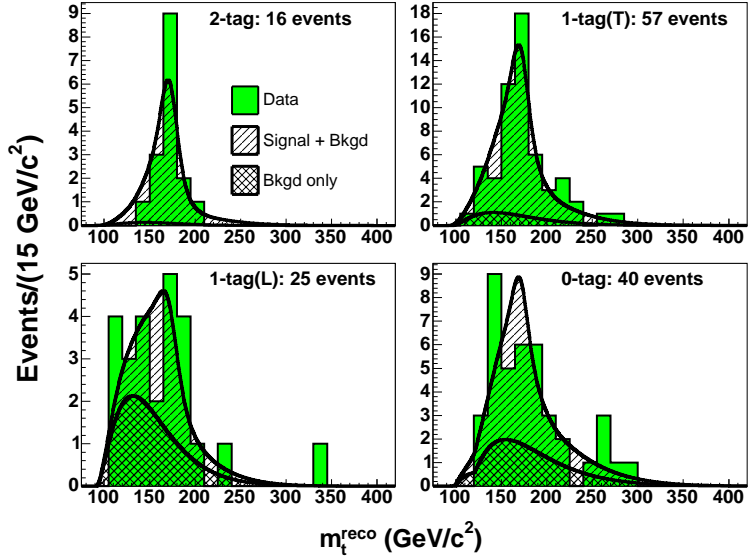}
 \caption{\it
      Reconstructed top mass in the four data subsamples (solid), with the best fits from the simulated templates overlaid (hatched). The contribution of background events is also shown (double-hatched). 
 \label{Fig1} }
\end{figure}
A two-dimensional likelihood fit yields both the measured top mass and the JES shift (in $\sigma$'s) from the a-priori estimate via independent means\cite{CDF_JES}. 
We get $M_{top} = 173.5 ^{+3.7}_{-3.6}$ (stat.) GeV/c$^2$, were 
the uncertainty is statistical and incorporates the JES contribution ($\sim 2.5$ GeV/c$^2$, to be compared to a $3.1$ GeV/c$^2$ contribution in the 1-D template method, where the hadronic $W$ decays are not used to constrain the JES).
The systematic uncertainties not included in the JES are small 
($1.3$ GeV/c$^2$), resulting in $M_{top} = 173.5 ^{+3.9}_{-3.8}$ GeV/c$^2$.

\subsection{Other measurements}
Matrix-element methods are used on both the $\ell + jets$ and the dilepton channels, and are proved to be very powerful statistically. 
We select events with exactly four or exactly two energetic jets for the 
$\ell + jets$ and the dilepton channels, respectively; this way, NLO 
effects are minimized when comparing the data with the LO matrix element for 
the $t\bar{t}$ production and decay.
With~63~$\ell + jets$ events containing at least one $b$-tagged jet 
each, a matrix-element analysis measures 
$M_{top} = 173.2 ^{+2.6}_{-2.4}$ (stat.) $\pm 3.2$ (syst.) $=$ $173.2 ^{+4.1}_{-4.0}$ GeV/c$^{2}$, where the JES contributes a 3 GeV/c$^2$ systematic uncertainty\cite{CDF_ljets_Mtop}.
In the dilepton channel, 33 events reconstructed in 340 pb${-1}$ of data 
yield $M_{top} = 165.3 \pm 6.3$ (stat.) $\pm 3.6$ (syst.) GeV/c$^{2}$ (with a 2.6 GeV/c$^2$ systematic uncertainty due to the JES)\cite{CDF_DILcombo_Mtop}.

The most precise measurement with a template method in the dilepton channel 
comes from the Neutrino Weighting Algorithm; for each top mas hypothesis 
we integrate over all possible neutrino $\eta$'s and calculate the probability that 
the measured missing energy is matched. The most probable top mass from each 
event serves as input to the template methodology (Section\ref{Sec:Mtop_methods}).
With 45 ``lepton plus track'' events reconstructed in 360 pb$^{-1}$ of data, 
this method gives $M_{top} = 170.7 ^{+6.9}_{-6.5}$ (stat.) $\pm 4.6$ (syst.) GeV/c$^{2}$, where the JES contributes a 3.4 GeV/c$^2$ systematic uncertainty\cite{CDF_DILcombo_Mtop}. 
Combining the dilepton measurements at CDF we get 
$M_{top}~=~167.9~\pm~5~2$~(stat.)~$\pm 3.7$~(syst.)~GeV/c$^{2}$\cite{CDF_DILcombo_Mtop}.

\section{Summary and outlook}
The top mass measurement is entering a precision phase. 
CDF has provided the single best measurement to date ($M_{top} = 173. 5^{+3.9}_{-3.8}$ GeV/c$^2$) by using a 2-dimensional template method which constraints the jet energy scale by using the dijet mass from the in-situ hadronic $W$ decays in
lepton plus jets $t\bar{t}$ candidate events.
The jet energy scale is the biggest contributor to the $M_{top}$ systematic 
uncertainty: $\sim 2.5$ GeV/c$^2$ with the 320 pb$^{-1}$ data sample of this 
measurement, 
but it's expected to contribute $\sim 1.5$ (1) GeV/c$^2$ when 2 (4) fb$^{-1}$ 
are collected and analyzed.

By using exclusive datasets and combining the best measurement from each 
channel, a preliminary Tevatron average
($M_{top} = 172.7 \pm 2.9$ GeV/c$^2$) was obtained in the summer 
of 2005 \cite{TeV_Mtop_2005}.
This is already a $1.7\%$ measurement.




\begin{thebibliography}{99}
%
\bibitem{LEPEW2005} LEP ElectroWeak Working Group, {\tt http://lepewwg.web.cern.ch/LEPEWWG/}, CERN-PH-EP/2005-051 and hep-ex/0511027, to be published (2005).
%
\bibitem{top_discovery} F.~Abe {\it et al.} (CDF Collaboration), Phys. Rev. Lett. {\bf 74}, 2626 (1995); S.~Abachi {\it et al.} (D{\O} Collaboration),  Phys. Rev. Lett. {\bf 74}, 2632 (1995).
%
\bibitem{Mtop_World_Run1} CDF and D{\O} Collaborations, hep-ex/0404010 (2004).
%
\bibitem{CDF_ljets_Mtop} A.~Abulencia {\it et al.} (CDF Collaboration), FERMILAB-PUB-05-472-E, submitted to Phys. Rev. {\bf D}, hep-ex/0510048 (2005);
A. Abulencia {\it et al} (CDF Collaboration), FERMILAB-PUB-05-474-E, submitted to Phys. Rev. Lett., hep-ex/0510049 (2005).
%
%
\bibitem{CDF_DIL-LTRK_xsec} D. Acosta {\it et al.} (CDF Collaboration), Phys. Rev. Lett. {\bf 93}, 142001 (2004).
%
\bibitem{CDF_JES} A. Bhatti {\it et al.}, submitted to Nucl. Instr. Meth. A, hep-ex/0510047 (2005).
%
\bibitem{CDF_DILcombo_Mtop} A. Abulencia {\it et al.} (CDF Collaboration), FERMILAB-PUB-05-551-E, submitted to Phys. Rev. Lett., hep-ex/0512070 (2005); 
A. Abulencia {\it et al} (CDF Collaboration), FERMILAB-PUB-06-019-E, submitted to Phys. Rev. {\bf D} hep-ex/0602008 (2006).
%
%
\bibitem{TeV_Mtop_2005} CDF Collaboration, D{\O} Collaboration and the Tevatron Electroweak Working Group, FERMILAB-TM-2323-E, hep-ex/0507091 (2005).
%
\end{thebibliography}
\end{document}